\documentclass[useAMS]{mn2e}
\usepackage{graphicx}

\title[AGB Stars in the Phoenix Dwarf Galaxy]{Asymptotic Giant Branch Stars in the Phoenix Dwarf Galaxy}
\author[Menzies et al.]{John Menzies$^1$, Michael Feast$^2$, 
Patricia Whitelock$^{1,2,3}$, Enrico Olivier$^{1,4}$,
\newauthor Noriyuki Matsunaga$^5$ and Gary Da Costa$^6$\\
      $^1$ South African Astronomical Observatory, P.O.Box 9, 7935
           Observatory, South Africa.\\
      $^2$ Astronomy Department, University of Cape Town, 7701 Rondebosch,
           South Africa.\\
      $^3$ National Astrophysics and Space Science Programme, Department of
           Mathematics and Applied Mathematics,\\ 
           University of Cape Town, 7701 Rondebosch, South Africa.\\
      $^4$ Department of Physics, University of the Western Cape, Private 
           Bag X17, 7535 Bellville, South Africa. \\
      $^5$ Department of Astronomy, Kyoto University, Kitashirakawa 
           Oiwake-cho, Sakyo-ku, Kyoto, Kyoto 606-8502, Japan.\\
      $^6$ Research School of Astronomy and Astrophysics, Australian
           National University, Mount Stromlo Observatory, Weston Creek,\\ 
           ACT 2611, Australia.\\
}

\begin{document}
\maketitle
\begin{abstract}
$JHK_s$ near-infrared photometry of stars in the Phoenix dwarf galaxy is 
presented and discussed. Combining these data with the optical photometry of
Massey et al. allows a rather clean separation of field stars from Phoenix
members. The discovery of a Mira variable (P = 425 days), which is almost
certainly a carbon star, leads to an estimate of the distance modulus of
$23.10\pm 0.18$ that is consistent with other estimates and indicates the
existence of a significant population of age $\sim 2$ Gyr. The two carbon
stars of Da Costa have $M_{bol} = -3.8$ and are consistent with belonging to
a population of similar age; some other possible members of such a population 
are identified. A Da Costa non-carbon star is $\Delta K_s \sim
0.3$ mag brighter than these two carbon stars. It may be an AGB star of the
dominant old population.  The nature of other stars lying close to it in the
$K_s,(J-K_s)$ diagram needs studying.
\end{abstract}
\section{Introduction}
The present investigation of the
Phoenix dwarf galaxy is part of a programme to study local
group galaxies  using the Japanese - South African 1.4m
Infrared Survey Facility (IRSF) and {\sc sirius} three-channel camera 
(Nagashima et al. 1999, Nagayama et al. 2003) at SAAO Sutherland.

Phoenix is a member of the Local Group and the most distant of the Milky
Way's satellite galaxies (e.g. Grebel (1999) fig. 3).  It was discovered by
Schuster \& West (1976) who originally suggested it might be a globular
cluster; Canterna \& Flower (1977) established that it was a galaxy. Though
its overall properties are consistent with a classification as a dwarf
spheroidal, it also contains a relatively small young component and is thus
often referred to as a dIrr/dSph (e.g. Mateo 1998). It is associated with an
off-centre H{\sc i} cloud (Oosterloo, Da Costa \& Staveley-Smith 1996; Young
\& Lo 1997; St-Germain et al. 1999). The origin of this cloud is not clear,
although it may be formed from supernovae winds associated with the most
recent epoch of star formation in the galaxy (Young et al. 2007). Though
there have been a number of optical studies of Phoenix, this seems to be the
first to describe $JHK_s$ observations.

\section{Observations}
Images centred on Phoenix were obtained over a period of about 3 years. A
single observation comprises 10 dithered 30-s exposures which were reduced
by means of the standard {\sc sirius} pipeline (Nakajima private
communication). Normally, three such sets of frames were combined to give an
effective 900-s exposure in each of $J, H$ and $K_{s}$; when the seeing was
poor, we combined six sets for an 1800-s exposure. Standard stars from
Persson et al. (1998) were observed on each night and the results presented
here are on the natural system of the {\sc sirius} camera, but with the
zero-point of the Persson et al. standards. These magnitudes are expected to
be close to those on the 2MASS system (Kato et al. 2007). The field of view
is $7.8 \times 7.8$ arcmin, but this is reduced to $7.2
\times 7.2$ arcmin during the course of the reductions.  The scale is 0.45
arcsec $\rm pixel^{-1}$. According to Canterna \& Flower (1977) the optical
size of Phoenix is $7 \times 9$ arcmin. Thus our observations cover most of
the galaxy.

Table 1 contains our $JHK_s$ results for all single stars measured on the 
images of Phoenix obtained for this investigation, together with
positions allowing cross-identifications to the optical photometry of Massey
et al. (2007; henceforth M2007), our identification number (N)
which will be used in the text, and, in the last column, the $I$ magnitude 
derived from M2007. Mean $JHK_s$ magnitudes from all frames in each colour 
were used in compiling the table. The limiting magnitude is about 17.65 in 
$K_s$ where the typical internal error is 0.04 mag; typical errors in $J$ 
and $H$ are 0.03 mag or less. Table 3 contains individual observations and 
dates of the two red variables found in our work, which are discussed in 
section 4.

The $K_s$ and $J$ frames were compared visually to check for possible very
red AGB stars, but none was found redder than the Mira (see section 4), down
to $K_s \sim 17$ mag.

\begin{table}
\begin{center}
\caption{Positions and IR photometry for all single stars measured on the 
images of Phoenix obtained for this investigation.}
\begin{tabular}{llrrrrrl}
 \multicolumn{1}{c}{RA} & \multicolumn{1}{c}{Dec} & N & $K_s$ & J-H & H-Ks & J-Ks & \multicolumn{1}{c}{I}\\
\multicolumn{2}{c}{(J2000.0)}&&&&&&\\
\hline
27.70577 &  -44.47683 &   50 &  16.93 &   0.65 &   0.04 &   0.68 &  18.34 \\
27.70883 &  -44.43503 &  122 &  17.54 &   0.71 &   0.08 &   0.78 &  19.16 \\
27.71133 &  -44.42600 &  129 &  17.46 &   0.69 &   0.07 &   0.76 &  19.11 \\
27.71861 &  -44.50923 &    8 &  15.36 &   0.71 &   0.17 &   0.88 &  17.24  \\
27.72032 &  -44.48024 &   49 &  16.69 &   0.36 &  -0.07 &   0.29 &  17.37  \\
27.72356 &  -44.46934 &   88 &  17.11 &   0.76 &   0.11 &   0.87 &  18.98 \\
27.72534 &  -44.49939 &   26 &  15.79 &   0.36 &   0.02 &   0.39 & $-$ \\
27.73110 &  -44.44578 &  107 &  17.36 &   0.82 &   0.06 &   0.88 &  19.21  \\
27.73291 &  -44.41672 &   13 &  15.19 &   0.64 &   0.07 &   0.70 &  16.69 \\
27.73303 &  -44.42099 &  131 &  17.55 &   0.75 &   0.04 &   0.80 &  19.26 \\
27.73541 &  -44.43295 &  123 &  17.39 &   0.81 &   0.13 &   0.94 &  19.26  \\
27.73560 &  -44.48583 &   81 &  17.03 &   0.59 &   0.18 &   0.77 &  19.26  \\
27.73603 &  -44.47464 &   31 &  15.69 &   0.65 &   0.17 &   0.82 &  17.73 \\
27.73785 &  -44.43591 &  121 &  17.28 &   0.57 &   0.22 &   0.79 &  $-$ \\
27.74408 &  -44.41630 &  134 &  17.31 &   0.78 &   0.16 &   0.93 &  19.17 \\
27.74641 &  -44.44115 &    3 &  13.15 &   0.64 &   0.04 &   0.69 &  $-$ \\
27.74851 &  -44.48817 &   46 &  16.52 &   0.88 &   0.11 &   0.98 &  18.64  \\
27.75006 &  -44.44932 &    2 &  13.11 &   0.50 &   0.01 &   0.51 &  80.00 \\
27.75215 &  -44.42229 &  130 &  17.12 &   0.75 &   0.09 &   0.84 &  18.88  \\
27.75259 &  -44.46943 &   52 &  16.59 &   0.85 &   0.09 &   0.94 &  18.61 \\
27.75265 &  -44.44532 &  108 &  17.46 &   0.66 &   0.03 &   0.69 &  18.89 \\
27.75280 &  -44.44059 &  120 &  17.44 &   0.81 &   0.12 &   0.93 &  19.52 \\
27.75411 &  -44.47115 &   51 &  15.03 &   1.19 &   0.89 &   2.08 &  $-$ \\
27.75417 &  -44.44467 &  112 &  17.32 &   0.66 &   0.03 &   0.70 &  18.78 \\
27.75427 &  -44.45249 &   12 &  15.39 &   0.83 &   0.12 &   0.95 &  17.42  \\
27.75427 &  -44.45694 &    7 &  14.49 &   0.44 &  -0.01 &   0.43 &  $-$ \\
27.75441 &  -44.44759 &   32 &  15.88 &   0.72 &   0.09 &   0.81 &  17.54  \\
27.75442 &  -44.44666 &  105 &  17.68 &   0.42 &   0.01 &   0.42 &  18.23 \\
27.75659 &  -44.40777 &  136 &  17.56 &   0.67 &   0.05 &   0.71 &  19.16 \\
27.75774 &  -44.45494 &   97 &  17.33 &   0.83 &   0.11 &   0.94 &  19.25 \\
27.75827 &  -44.44532 &   56 &  16.54 &   0.86 &   0.13 &   0.99 &  18.64  \\
27.76108 &  -44.41770 &  133 &  17.34 &   0.75 &   0.10 &   0.85 &  19.18  \\
27.76173 &  -44.44854 &   55 &  16.49 &   0.75 &   0.07 &   0.82 &  18.18 \\
27.76268 &  -44.44494 &  110 &  17.47 &   0.82 &   0.07 &   0.89 &  19.23 \\
27.76659 &  -44.48769 &   47 &  16.58 &   0.67 &   0.10 &   0.77 &  18.22  \\
27.76702 &  -44.44343 &  114 &  17.65 &   0.59 &   0.07 &   0.66 &  19.00  \\
27.76729 &  -44.45500 &   96 &  17.44 &   0.75 &   0.10 &   0.84 &  19.24  \\
27.76815 &  -44.45673 &   94 &  17.60 &   0.80 &   0.12 &   0.92 &  19.40  \\
27.76831 &  -44.48687 &   48 &  16.64 &   0.62 &   0.17 &   0.79 &  18.38 \\
27.76916 &  -44.42640 &    5 &  13.95 &   0.38 &  -0.00 &   0.38 &  $-$ \\
27.76964 &  -44.45859 &    6 &  14.32 &   0.59 &   0.17 &   0.76 &  16.11 \\
27.77019 &  -44.48075 &   83 &  17.20 &   0.74 &   0.10 &   0.84 &  19.00 \\
27.77352 &  -44.48684 &   80 &  17.30 &   0.82 &   0.13 &   0.94 &  19.13 \\
27.77360 &  -44.43816 &   33 &  15.94 &   0.87 &   0.27 &   1.14 &  18.24 \\
27.77402 &  -44.49752 &   10 &  15.47 &   0.36 &  -0.00 &   0.36 &  16.24 \\
27.77405 &  -44.45512 &   53 &  16.24 &   0.68 &   0.14 &   0.82 &  18.05 \\
27.77557 &  -44.43107 &  125 &  17.56 &   0.81 &   0.12 &   0.93 &  19.43 \\
27.77731 &  -44.44379 &  113 &  17.46 &   0.75 &   0.10 &   0.85 &  19.23  \\
27.77740 &  -44.43351 &   60 &  16.73 &   0.82 &   0.14 &   0.96 &  18.69 \\
27.77798 &  -44.45670 &   93 &  17.43 &   0.80 &   0.13 &   0.93 &  19.36 \\
27.77821 &  -44.50607 &   71 &  17.22 &   0.78 &   0.15 &   0.92 &  19.08  \\
27.77984 &  -44.44510 &  109 &  17.48 &   0.82 &   0.10 &   0.91 &  19.36 \\
27.78000 &  -44.45448 &   98 &  17.38 &   0.83 &   0.13 &   0.96 &  19.29 \\
27.78031 &  -44.44961 &  104 &  17.05 &   0.74 &   0.08 &   0.83 &  18.82 \\
27.78081 &  -44.44227 &  116 &  16.83 &   0.87 &   0.36 &   1.23 &  $-$ \\
27.78131 &  -44.45070 &  103 &  17.50 &   0.74 &   0.14 &   0.88 &  19.25 \\
27.78158 &  -44.48930 &   78 &  17.09 &   0.67 &   0.11 &   0.78 &  18.60 \\
27.78170 &  -44.45373 &   99 &  17.62 &   0.84 &   0.10 &   0.94 &  19.51  \\
27.78273 &  -44.41102 &   14 &  14.72 &   0.41 &   0.03 &   0.44 &  15.25 \\
27.78391 &  -44.45143 &  101 &  17.29 &   0.83 &   0.09 &   0.92 &  19.20  \\
27.78412 &  -44.47465 &   86 &  17.46 &   0.77 &   0.09 &   0.86 &  19.26  \\
27.78465 &  -44.47567 &   84 &  17.53 &   0.76 &   0.12 &   0.88 &  19.39 \\
\hline
\end{tabular}
\end{center}
\end{table}

\setcounter{table}{0}
\begin{table}
\begin{center}
\caption{Continued}
\begin{tabular}{llrrrrrl}
\multicolumn{1}{c}{RA} & \multicolumn{1}{c}{Dec} & N & $K_s$ & J-H & H-Ks & J-Ks & \multicolumn{1}{c}{I}\\
\multicolumn{2}{c}{(J2000.0)}&&&&&&\\
\hline
27.78534 &  -44.50643 &   41 &  16.59 &   0.82 &   0.17 &   1.00 &  18.68  \\
27.78607 &  -44.45109 &   54 &  16.80 &   0.68 &   0.08 &   0.76 &  18.35 \\
27.78643 &  -44.42723 &  128 &  17.48 &   0.69 &   0.23 &   0.91 &  19.48 \\
27.78651 &  -44.43156 &  124 &  17.29 &   0.81 &   0.11 &   0.93 &  19.23  \\
27.78669 &  -44.44216 &  117 &  17.75 &   0.72 &  -0.00 &   0.72 &  19.33 \\
27.78671 &  -44.42488 &   34 &  15.25 &   1.05 &   0.53 &   1.57 &  19.48 \\
27.78732 &  -44.47034 &   87 &  16.90 &   0.81 &   0.15 &   0.96 &  19.00 \\
27.78837 &  -44.48860 &    4 &  13.30 &   0.45 &   0.01 &   0.46 &  $-$ \\
27.79029 &  -44.47513 &   85 &  17.37 &   0.83 &   0.13 &   0.96 &  19.29  \\
27.79328 &  -44.42856 &  126 &  17.02 &   0.79 &   0.12 &   0.91 &  19.00 \\
27.79454 &  -44.51225 &   40 &  16.26 &   0.69 &   0.17 &   0.86 &  17.92  \\
27.79501 &  -44.41957 &   61 &  16.76 &   0.65 &   0.06 &   0.71 &  18.29 \\
27.79895 &  -44.45101 &  102 &  16.85 &   0.67 &   0.14 &   0.81 &  18.55  \\
27.80202 &  -44.45618 &   95 &  17.54 &   0.77 &   0.10 &   0.87 &  19.32 \\
27.80302 &  -44.40294 &   35 &  16.01 &   0.58 &   0.09 &   0.67 &  17.49  \\
27.80355 &  -44.40904 &  135 &  17.53 &   0.76 &   0.13 &   0.89 &  19.41 \\
27.80573 &  -44.46179 &   91 &  17.43 &   0.78 &   0.02 &   0.81 &  19.13 \\
27.80617 &  -44.50018 &   72 &  17.55 &   0.67 &   0.07 &   0.74 &  19.67 \\
27.80866 &  -44.46093 &   92 &  17.57 &   0.83 &   0.11 &   0.93 &  19.43 \\
27.80904 &  -44.44226 &   57 &  16.56 &   0.87 &   0.16 &   1.04 &  18.78  \\
27.81347 &  -44.50706 &   70 &  17.43 &   0.78 &   0.14 &   0.91 &  19.27 \\
27.81846 &  -44.40759 &  137 &  17.36 &   0.74 &   0.04 &   0.79 &  19.19  \\
27.81927 &  -44.51210 &   69 &  17.33 &   0.67 &   0.17 &   0.83 &  18.95 \\
27.82514 &  -44.39088 &   62 &  16.63 &   0.56 &   0.23 &   0.79 &  $-$ \\
27.82563 &  -44.45501 &   11 &  15.37 &   0.68 &   0.12 &   0.81 &  17.15  \\
27.82798 &  -44.46254 &   90 &  17.55 &   0.78 &   0.08 &   0.85 &  19.33  \\
27.83021 &  -44.44322 &  115 &  17.63 &   0.71 &   0.05 &   0.76 &  19.28 \\
27.83413 &  -44.39202 &  141 &  17.55 &   0.59 &   0.25 &   0.84 &  19.45 \\
27.83445 &  -44.48040 &   30 &  15.66 &   0.71 &   0.07 &   0.78 &  17.28 \\
27.83794 &  -44.44127 &  118 &  17.53 &   0.79 &   0.14 &   0.93 &  19.38  \\
27.83868 &  -44.46736 &   89 &  17.25 &   0.61 &   0.18 &   0.79 &  18.97  \\
27.84202 &  -44.45187 &  100 &  17.59 &   0.40 &   0.12 &   0.52 &  18.15  \\
27.84331 &  -44.42792 &  127 &  17.60 &   0.77 &   0.19 &   0.95 &  19.48 \\
\hline
\end{tabular}
\end{center}
\end{table}

\section{Colour-Magnitude and Colour-Colour Diagrams}

Fig.~1 shows the $K_s,(J-K_s)$ diagram and Fig.~2 the $(J-H),(H-K_s)$
diagram for all single stars measured on the images centred on Phoenix that 
were obtained for this investigation. The photometry reduction program
considered another five objects as double, and these were not plotted as the
individual magnitudes were too uncertain.  In discussing these figures we
assume a distance modulus of 23.1 mag for Phoenix (see section 4).  The blue
stars in a vertical sequence with $(J-K_s) \sim 0.4$ in Fig.~1 are almost
certainly field stars. This can be seen, for instance, by comparing with the
similar figures in Menzies et al. (2002; henceforth JWM2002) for Leo~I,
obtained with the same instrumental arrangement. There are somewhat fewer of
these stars in the Phoenix field. This is probably due to the higher
galactic latitude ($b = -69$ (Phoenix); $b = +49$ (Leo~I)). Other stars
which are likely to be field stars (see below) are also marked as
asterisk-shaped symbols.  In Fig 2. the clear separation of many of the likely 
field stars from the members is apparent. There is some similarity of the
distribution of points in this diagram with that for Leo~I (JWM2002);
there is a clump of stars with $(H-K_s) < 0.3$ as in Leo~I, 
and a few redder ones that in Leo~I are all carbon stars and mostly variable.
A comparison between Phoenix and Leo~I is made below (see section 3).
The two carbon stars discovered by Da Costa (1994) are
also marked on these figures.

\begin{figure} \includegraphics[width=8.5cm]{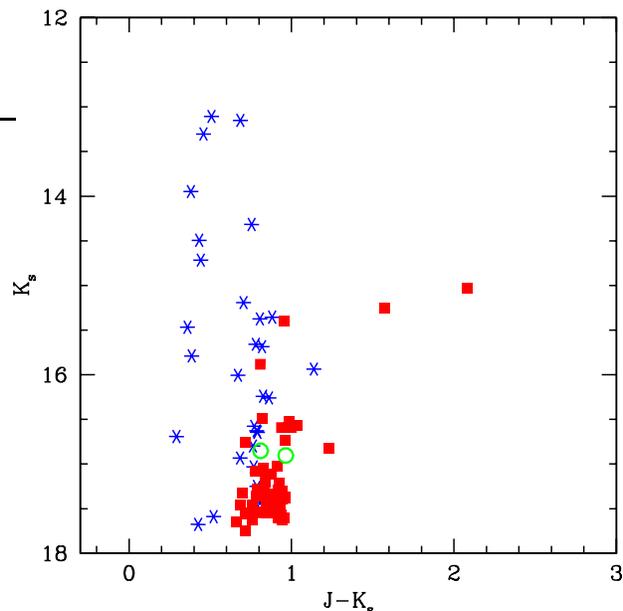}
\caption{Colour-magnitude diagram for the field centred on Phoenix.
Asterisk symbols are probable field stars (i.e. they lie above the dashed
line in Fig\,3(a)). Filled squares are probable
members, while the two known carbon stars are shown as open circles. }
\label{fig_kjk} 
\end{figure}

\begin{figure}
\includegraphics[width=8.5cm]{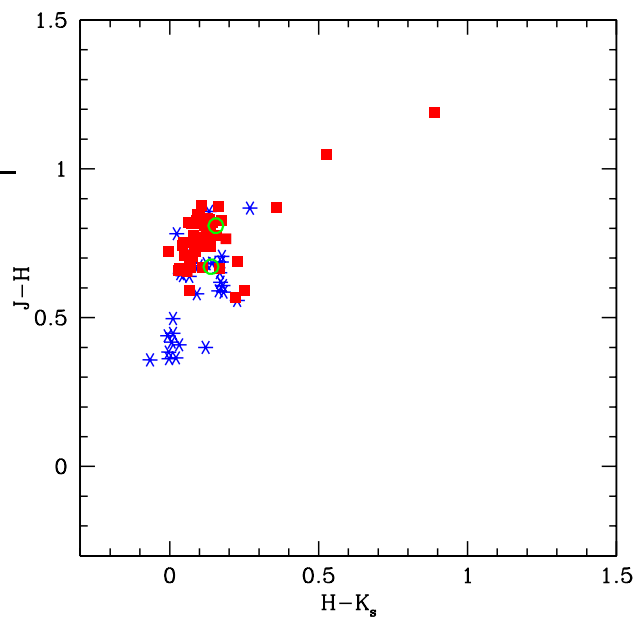}
\caption{$(J-H),(H-K_s)$ two-colour diagram for the Phoenix field. 
Symbols as for Fig.~1. }
\label{fig_jhhk}
\end{figure}

The interpretation of the colour-magnitude diagram and the elimination of
likely foreground stars is helped considerably by combining our data with
the optical photometry of Massey et al. (2007). This is
particularly important in establishing the AGB population of the galaxy.
Held et al. (1999) have suggested (see their fig. 11) that there is a
significant population of AGB stars in Phoenix with $I < 19.5$ and
$(B-I) > 3.0$, but the separation from field stars is difficult
(cf.\ Martinez-Delgado et al.\ 1999).

Fig 3(a) is a $(V-R),(B-V)$ diagram for objects with $I < 19.5$ and quoted
uncertainties in both co-ordinates less than 0.1 mag from the observations
of M2007. These cover an area of 34 x 34 arcmin centred on Phoenix.
Since this area is much larger than the galaxy itself, the bulk of the 
stars plotted are field objects. In particular, the heavily populated
areas in this diagram are likely to contain a high proportion
of field stars.
Especially at the redder colours the stars in this diagram divide
rather clearly into two groups (presumably giants and dwarfs). The dashed
line, extrapolated to bluer colours, approximately marks this division.

\begin{table}
\begin{center}
\caption{Stars in common with Da Costa (1994).} 
\begin{tabular}{lrrrrrr}
   DaCosta   &  vdRDK  &  This & Spectroscopy\\
\hline
C1 & 481 &87 & carbon  \\
C2 &     & 52 & not carbon \\
C3 & 391?& 88 & not carbon \\
C4 & 166 & 102 & carbon  \\
C5 &     & 106 & not carbon \\
\hline
\end{tabular}
\end{center}
Note: Stars originally selected by M Irwin. vdRDK: van de Rydt et al. (1991),
\end{table}

In Fig 3(b) the dashed line from Fig 3(a) is repeated and stars in common
between out survey and M2007 are plotted.  The curves show the loci of normal 
giants and dwarfs.  Using the division into
two sequences and the density of the M2007 points in Fig 3(a), we divide
our stars into probable field stars (asterisks) and probable
Phoenix members (squares). The differing distribution of the stars in
the whole M2007 sample and those in common with our survey, strongly
suggests that the stars below the dashed line (the likely giant region)
have a high probability of being Phoenix members and we have taken
them as such. Note particularly the concentration of stars in 
common, below the dotted line and with $\sim 1.4 < (B-V) < \sim 1.55$,
strongly indicating membership.
The Bahcall-Soneira model (Bahcall \& Soneira 1980)
predicts 24 field stars with $V < 20$ for a field of our size at $b = 90$
and 32 at $b = 50$, $l = 270$ (Phoenix has $b = -69$, $l =272$). In view of
the small number statistics the number of stars rejected here (35) seems to
be of the correct order.  Fig.~4 is an $I,V-I$ diagram of stars in common
with M2007 with field stars and members distinguished. The distribution of
probable field stars in this figure supports their classification as such. 

\begin{figure*}
\includegraphics[width=17.0cm]{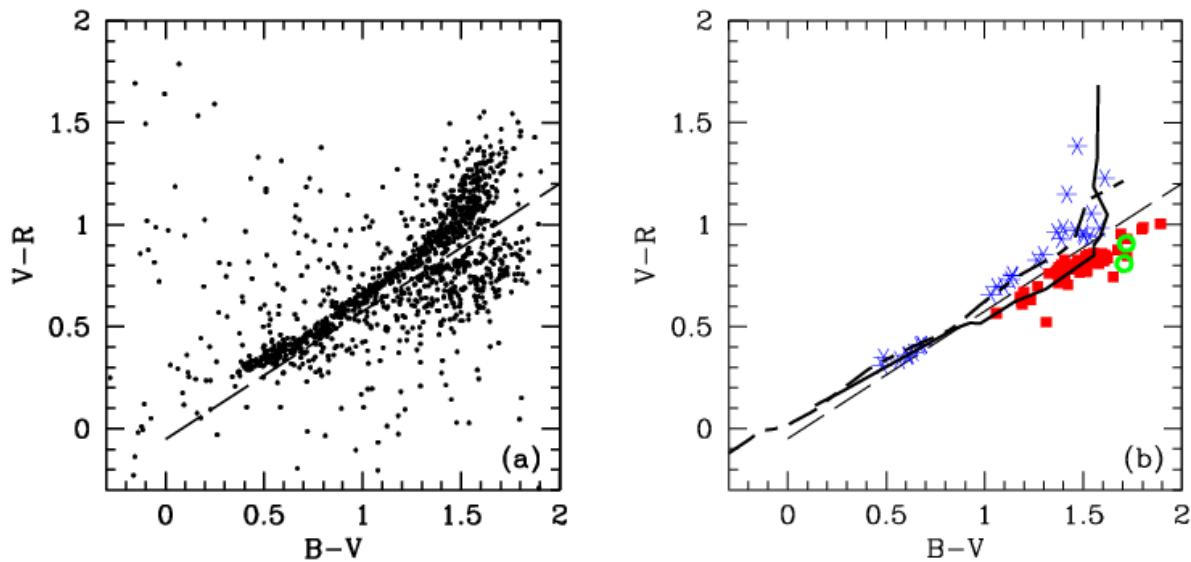}
\caption {(a)$(V-R),(B-V)$ two-colour diagram for the Phoenix field.
Small squares represent stars from M2007 (see text).
The long-dashed line shows our proposed separation between
field and Phoenix stars. (b)Stars in our survey in common with M2007 are 
plotted in a $(V-R),(B-V)$ two-colour diagram, with asterisk symbols being 
probable field stars and filled
squares probable Phoenix members; the two carbon stars are indicated by open
circles. Dash-dot and continuous
lines represent standard dwarf and giant sequences (derived from Bessell
(1990)).}
\end{figure*}

There is a sparse population of young, blue, stars in Phoenix. This has an
upper brightness limit of about $V= 19$ (M2007 fig. 20). Since the $V-K_s$
colours of these stars must be within a few tenths of a magnitude of zero,
they will be too faint to be in our survey.

\begin{figure}
\includegraphics[width=8.5cm]{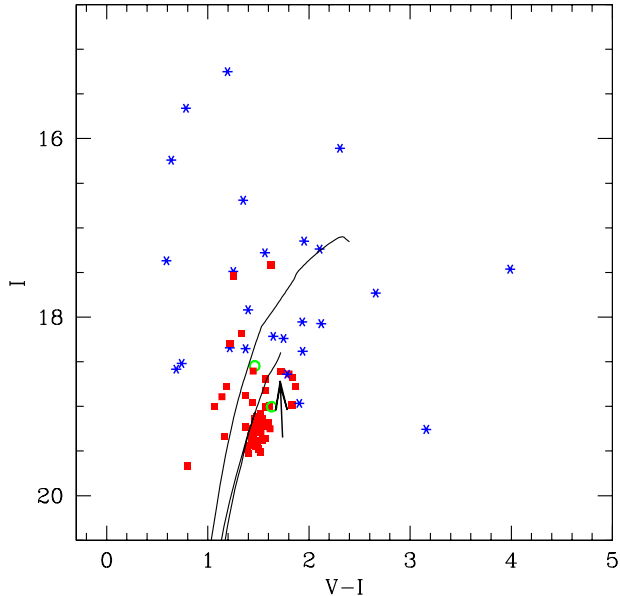}
\caption{$I,(V-I)$ colour-magnitude diagram for stars in Phoenix measured in 
this programme. Symbols as in Fig.~1. The curves illustrate isochrones from 
Girardi et al. (2000, 2002) for two populations, one with age 14\,Gyr and metallicity
z=0.001 and the other with age 1\,Gyr and z=0.002. The two red variables, with
$(J-K_s)>1.4$, do not have $I,(V-I)$ photometry and thus cannot be included in 
the figure. Star C2 (our number 52) is marked with an arrow.}
\end{figure}

Figs.~5 and 6 are $K_s,(J-K_s)$ and $(J-H),(H-K_s)$ plots for Phoenix with
the likely field stars removed. In Figs.~4 and 5 we show isochrones (RGB and
AGB) for z = 0.001, age 14 Gyr and z = 0.002, age 1 Gyr from Girardi et al.
(2000, 2002). The RGB tip (TRGB) for the 14 Gyr model occurs at $K=17.17$ and
$I=19.06$ (i.e. very close to the values observed, 
as will be discussed below), while that for the 1 Gyr model is at $K=19.87$ 
and $I=21.17$\footnote{As will be discussed in section 4, the reddening of 
the Phoenix stars is small enough to be neglected in these comparisons with 
iscochrones.}.  Note that the isochrones provide only a
qualitative illustration of the populations which might be present and they
suggest that the members of the dwarf galaxy are on the RGB and AGB of a
population with a large range of ages. These late stages of stellar
evolution are not well understood and models by different authors provide
significantly different tracks, e.g. AGB isochrones from Pietrinferni et al.
(2004) terminate several magnitudes fainter than those illustrated.
Furthermore, there are as yet no examples of AGB models which provide a good
fit to observations over a range of wavelengths.

\begin{figure}
\includegraphics[width=8.5cm]{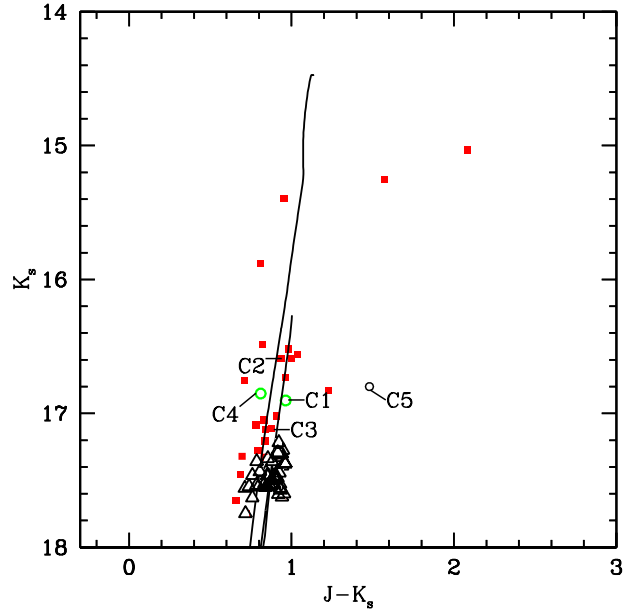}
\caption{$K_s,(J-K_s)$ colour-magnitude diagram for Phoenix showing only
the probable members. 
Open triangles indicate stars with $I > 19.05$, corresponding to the 
majority old population RGB in Phoenix. Other symbols are as in Fig.~1 and 
isochrones are as in Fig.~4.}
\end{figure}
\begin{figure}
\includegraphics[width=8.5cm]{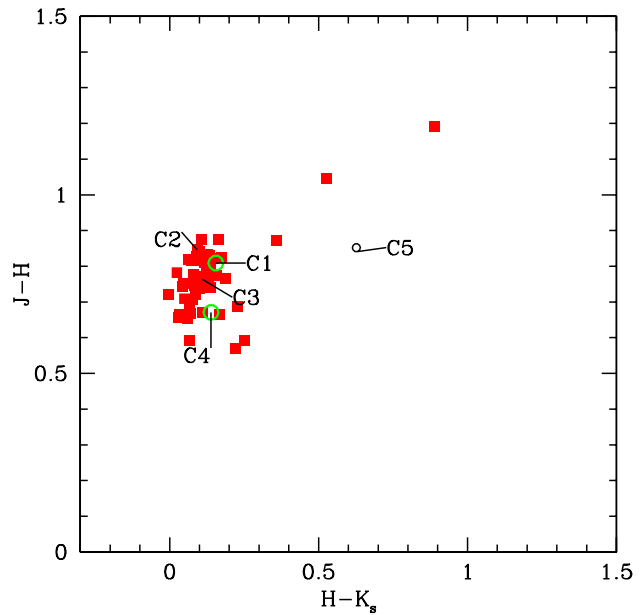}
\caption{$(J-H),(H-K_s)$ two-colour diagram for Phoenix, without the probable field
stars. Symbols as in Fig.~1.}
\end{figure}

M2007 suggest that Phoenix may contain a population of yellow supergiants,
though they recognize that it is difficult in their work to distinguish
galaxy members from foreground stars. Since their yellow supergiant sequence
extends up to $V$ of about 15 mag and since $V-K_s$ may well be significant (1
or greater), it seems possible that the bright star in Fig.~5 with $J-K_s$ of
about 0.9 and $K_s$ of about 15.3 may be such a star.

The absolute magnitude of the TRGB in $K_s$ depends on metallicity and age. 
Estimates of the metallicity of the old population of Phoenix ranging from
--1.37 to --1.8 have been given (Gallart et al.\ 2004; Holtzman et al.\
2000; Held et al.\ 1999).  If the main population of Phoenix is of globular
cluster age and has a metallicity of $\sim -1.3$ we would expect the TRGB to
be at $M_{K} \sim -5.8$ (Salaris \& Girardi 2005), corresponding to $K_s =
17.3$ in Phoenix. We would then identify the near vertical, sequence of
stars with $(J-K_s) \sim 0.9$ and fainter than $K_s \sim 17.1$ as mainly due
to this population. 
Such a population will not produce normal, intrinsic, carbon stars, which are 
expected to belong to an intermediate age population. The possibility that 
these objects are extrinsic carbon stars seems remote, especially since they 
lie above the TRGB of even the old population (see section 4).

  Our identification of the TRGB of the main population at
$K_{s} \sim 17.1$ is entirely consistent with data at other wavelengths.
The presence of a TRGB in the range $I = 19.25$ to 19.00
was clearly established in the work of van de Rydt et al. (1991),
Martinez-Delgado et al. (1999), Held et al. (1999),
and Holtzman et al. (2000). Table A1 of our appendix shows that
all our probable AGB stars have $I$ magnitudes brighter than this.
(The only one close to the tip is one of the carbon stars).
A TRGB near $I = 19.0$ is also evident in Fig.~4 and the stars there
are those showing a tip at $K_{s} \sim 17.0$ in Fig.~5. 
One can also compare the results for Phoenix with RGB predictions.
The recent review by Bellazzini (2007), his fig.~5, leads to
a predicted TRGB at $K \sim 16.94$ for a colour at the tip of
$(J-K) \sim 0.9$ as in our case and for our adopted distance modulus.
As regards $I$, his fig.~4 (or his eq.~2, corrected for errors in sign)
leads to a TRGB of $\sim 19.08$ at our adopted distance and with
$(V-I)_{o} = 1.48$ at the tip (Held et al. 1999). Evidently there is
good consistency between the results from the TRGB at $I$ and that at
$K_{s}$.

In addition to finding two carbon stars, Da Costa (1994) obtained spectra of
three other stars in which he found no evidence of carbon-star features; we
presume these to be oxygen-rich. They are listed in Table 3 with their
identifications from van de Rydt et al. (1991) and our numbers. The
non-carbon stars are also marked in Figs.~5 and 6. C5 is a double, possibly
of field dwarfs.  C2 and C3 are among our presumed members. The position of
C3 in Fig.~5, immediately below the two C stars, suggests that it may be an
intermediate age star marking a lower limit to carbon star formation or it
may be a member of the old population and near its TRGB.

In Fig.~5 there are a number of stars fainter than $K_s \sim 17.2$ which are 
plotted as squares because they have $I$ magnitudes brighter than the TRGB (they are mainly in the range $I \sim 18.5$ to 19.0. In both Figs.~4 and 5 these 
stars lie to the blue of the main concentrations. The most likely explanation 
of these stars is that they are AGB stars of an intermediate age population and
 may well be coeval with the carbon stars.

The seven stars immediately above the two  carbon stars (and slightly
redder) in Fig.~5 constitute an interesting problem. Table A1 in the appendix 
lists optical and infrared data for these stars and for the two spectroscopic
carbon stars. Evidently the seven stars have colours rather similar to the
known carbon stars and on these grounds would be strong C star candidates.
This, together with their position immediately above the C stars in Fig.~6,
would be entirely in accord with expectation (see for instance the
distribution of C stars in the Leo~I $K_s,(J-K_s)$ diagram (JWM2002). However,
one of these stars, Da Costa C2, our No. 52, is not a spectroscopic carbon
star.  The nature of this star and possibly of the other six stars in this
group remains to be determined.

Table A2 in the appendix lists standard infrared sequences for giants and
dwarfs from Bessell \& Brett (1988) that have been converted to the 2MASS
system (which is close to the IRSF system) using the relations in Carpenter
(2001). Table A3 contains optical data for extreme subdwarfs listed by Gizis
(1997) together with 2MASS data for these stars. Comparison of these tables
with the data for our seven stars shows the following: their $J-H$ and $J-K_s$
are too red for them to be normal dwarfs. From their optical colours some,
or all, of them (including C2 = 52) could be extreme subdwarfs (compare
their positions in a plot such as fig. 9 of Gizis).  However, their infrared
colours are too red for such an assignment, being reasonably similar to
those of late type giants. On the other hand, several have suspiciously red
$B-V$ values both for normal giants and for metal-poor (globular cluster
type) giants. Furthermore, as late-type giants they would be very distant
(extragalactic) and thus strong Phoenix candidates.  Comparison with the
isochrones shown in Figs.~4 and 5 suggests that C2 is best interpreted as an
AGB star of the very old population. Evidently, spectral type and radial
velocity data are required for the other stars in this group to determine
whether or not they are similar to C2 or whether they are carbon stars.

A comparison of Figs.~5 and  6 with the $K_s,(J-K_s)$ and $(J-H),(H-K_s)$
diagrams of the dwarf spheroidal Leo~I (JWM2002) is of interest. The
reddening of both galaxies is small and the distance modulus of Phoenix is
$\sim 1.0$ mag greater than that of Leo~I. The faintest known carbon stars
in Leo~I are at $K_s \sim 15.5$ and are thus intrinsically brighter by $\sim
0.4$ mag than the Da Costa (1994) carbon stars in Phoenix. The two very red
and variable stars in Phoenix (see section 4) have $K_s \sim 15.2$ whilst
there are red variables in Leo~I with $K_s$ in the range 13.8-14.5 and so
roughly equivalent. However, there is a marked difference in the AGB between
the two galaxies. In Leo~I there is a well populated sequence of carbon
stars extending from about $K_s,(J-K_s),$ 15.5, 0.9 to 13.9, 1.7.  In
Phoenix, whilst the low-amplitude variable (star 34) lies near the upper end
of such a sequence at $K_s \sim 15.25$ and $(J-K_s) \sim 1.6$ there are no
stars in this sequence between it and the small clump of stars containing
C2.  The difference between Phoenix and Leo~I is even more marked, if, as
discussed above the stars in this clump are not C stars. These differences
are most likely related to the different star formation histories of these
dwarf galaxies.

\section{Variable Stars and Distance}
As Figs.~1 and 2 show there are two outstanding red stars in Phoenix.  We
find both stars to be variable.  The $JHK_s$ observations are listed in Table
3 with the Julian Dates for the observations.  Star number 34
is a low amplitude variable.  No convincing period is evident in our
measures though it varies on a time scale of 200 to 300 days. The $JHK_s$
measurements for this star in Table 1 are simply the means of the individual
values in Table 3.

\begin{table}
\begin{center}
\caption{$JHK_s$ photometry for two red variables in Phoenix.}
\begin{tabular}{lrrrrrr}
   JD   &  $J$  &  $\sigma_J$ & $H$ & $\sigma_H$  &  $K_s$  & $\sigma_K$  \\
\hline
\multicolumn{7}{c}{Star 51}\\
2507.11817 & 17.979 & 0.024 & 16.610 & 0.014 & 15.562 & 0.013 \\
2813.14560 & 16.961 & 0.008 & 15.717 & 0.004 & 14.860 & 0.005 \\
2881.02012 & 17.652 & 0.014 & 16.237 & 0.007 & 15.190 & 0.005 \\
2962.85237 & 18.070 & 0.030 & 16.655 & 0.009 & 15.498 & 0.008 \\
3010.80022 & 18.051 & 0.039 & 16.618 & 0.010 & 15.410 & 0.008 \\
3173.12370 & 16.577 & 0.009 & 15.479 & 0.008 & 14.718 & 0.009 \\
3256.06266 & 16.981 & 0.010 & 15.763 & 0.010 & 14.848 & 0.010 \\
3260.06350 & 16.993 & 0.009 & 15.798 & 0.009 & 14.879 & 0.005 \\
3292.92197 & 17.271 & 0.014 & 16.020 & 0.010 & 15.059 & 0.010 \\
3349.86790 & 17.565 & 0.020 & 16.315 & 0.009 & 15.242 & 0.011 \\
3352.82444 & 17.579 & 0.017 & 16.308 & 0.009 & 15.302 & 0.009 \\
3440.73494 & 17.204 & 0.026 & 16.044 & 0.016 & 15.143 & 0.014 \\
3531.16565 & 16.284 & 0.009 & 15.351 & 0.007 & 14.711 & 0.008 \\
3612.08834 & 16.372 & 0.015 & 15.308 & 0.010 & 14.692 & 0.020 \\
3615.01896 & 16.412 & 0.010 & 15.361 & 0.009 & 14.684 & 0.009 \\
\hline
\multicolumn{7}{c}{Star 34}\\
2507.11817 & 16.881 & 0.010 & 15.819 & 0.008 & 15.286 & 0.011 \\
2813.14560 & 17.033 & 0.012 & 15.942 & 0.004 & 15.345 & 0.005 \\
2881.02012 & 16.887 & 0.010 & 15.846 & 0.005 & 15.265 & 0.006 \\
2962.85237 & 16.823 & 0.010 & 15.807 & 0.007 & 15.259 & 0.009 \\
3010.80022 & 16.805 & 0.014 & 15.749 & 0.007 & 15.221 & 0.008 \\
3173.12370 & 16.916 & 0.012 & 15.839 & 0.009 & 15.272 & 0.009 \\
3256.06266 & 16.833 & 0.010 & 15.777 & 0.008 & 15.269 & 0.009 \\
3260.06350 & 16.822 & 0.009 & 15.782 & 0.010 & 15.255 & 0.007 \\
3292.92197 & 16.720 & 0.009 & 15.698 & 0.009 & 15.173 & 0.009 \\
3349.86790 & 16.757 & 0.009 & 15.738 & 0.008 & 15.205 & 0.013 \\
3352.82444 & 16.800 & 0.010 & 15.721 & 0.007 & 15.202 & 0.009 \\
3440.73494 & 16.889 & 0.019 & 15.816 & 0.014 & 15.287 & 0.015 \\
3531.16565 & 16.617 & 0.013 & 15.653 & 0.010 & 15.162 & 0.010 \\
3612.08834 & 16.814 & 0.020 & 15.769 & 0.015 & 15.214 & 0.026 \\
3615.01896 & 16.699 & 0.009 & 15.673 & 0.009 & 15.199 & 0.009 \\
\hline
\end{tabular}
\end{center}
\end{table}

Star 51 is a large amplitude (Mira) variable with P = $425 \pm 25$ days,
Fourier mean magnitudes of $ J = 17.11, H = 15.92, K_s = 15.03$ and $\Delta K_s
= 0.76, \Delta J = 1.52$.  Light curves are shown in Fig.~7.  The $J$ light
curve clearly shows that the star was brightening on a long time scale during the
time of our observations.  This is typical behaviour for a carbon Mira
(e.g. Whitelock et al. 2006). Such stars are obscured from time to time owing
to the ejection of material into the line of sight.

\begin{figure}
\includegraphics[width=8.5cm]{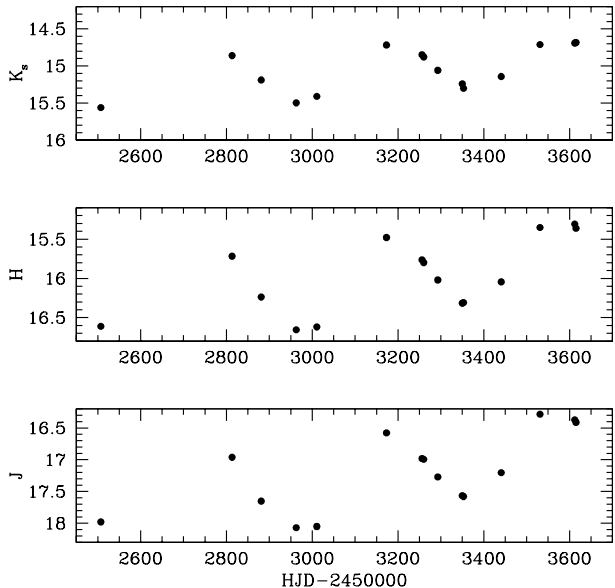}
\caption{The $JHK_s$ light curves of the Mira variable (star 51 in Table~1).}
\end{figure}

Previous authors (van de Rydt et al. 1991, Martinez-Delgado et al. 
1999, Held et al. 1999, Holtzman et al. 2000) have adopted a reddening
of $E(B-V) = 0.02$ mag from the work of Burstein \& Heiles (1982)
and we adopt this value. Whilst there is undoubtedly some uncertainty
in this, the reddening at the galactic latitude of Phoenix
($b = -69$) must be small and any reasonable change will have
little or no effect on our discussion. Then, converting the photometry
to the SAAO system using the relations of Carpenter (2001) we
obtain for the Mira (star 51) $K_{0} = 15.01, (H-K)_{0} = 0.90, (J-K)_{0} =
2.20$ and $m_{bol} = 18.51$. Here the bolometric correction to $K$ was
derived from the relation given by Whitelock et al. (2006, their equation
10), which is derived for carbon stars. We assume both this star and the
other variable are carbon stars in view of their red colours and the
presence of other carbon stars in the galaxy.  

The Mira (star 51) is the candidate long period variable ID 11263 of Gallart
et al. (2004). For this star their mean results are $I = 18.90, (B-V) =
4.38, V-I = 2.29$. It is the reddest likely member in their $I -(V-I)$
diagram. Since it lies on a reasonable extension of the RGB/AGB and not
fainter as it would be if it suffered from strong circumstellar extinction,
the very red $(B-V)$ must be mainly intrinsic to the star and is consistent
with its being a carbon star rather than a highly reddened oxygen-rich Mira. 
Comparison with figs.~7 and 8 of Whitelock et al. (2006) shows that in
$(J-H),(H-K_s)$ diagrams the two Phoenix variables lie close to  Galactic
and LMC carbon-rich Miras and SR variables. In an $(H-K_s),\log P$ diagram
(Whitelock et al. 2006, fig.~11) the Phoenix Mira lies in the region
occupied by Galactic carbon Miras and bluer than the known LMC carbon Miras,
except for two LMC stars believed to be undergoing hot-bottom burning (HBB). 
It is unlikely that it is an HBB star (see below). Whether this difference
from the LMC carbon stars is significant or not depends at least partly on
whether Phoenix contains redder (i.e. more dust enshrouded) stars which
might be below our detectable brightness limit.

Of the six candidate LPVs identified by Gallart et al. (2004) only three
fall in the area we surveyed. One of these is the Mira discussed above the
other two are our numbers 88 (their ID 11200) and 102 (their ID 8563). The
photometry of these two stars has marginally larger standard deviations in
all colours than stars of corresponding brightness, but there is no clear
periodicity in either. Note that star 102 is a Carbon star (C4 of Da Costa)
and the star 88 is NOT a Carbon star (C3 of Da Costa). No other stars brighter
than $K_s = 17.4$ show any convincing evidence of variability with amplitude 
greater than 0.1mag.

An estimate of the distance modulus of Phoenix can be made using
absolute magnitudes in either $K$ or $M_{bol}$ derived from period 
luminosity relations. In the case of $K$ we use the relation:
\begin{equation}
M_{K} = -3.30\log P + 0.59.
\end{equation}
The slope of this relation was derived from carbon Miras in the LMC by
Feast et al. (1989), and we have assumed an LMC distance modulus
of $18.39 \pm 0.05$ (van Leeuwen et al. 2007). For $M_{bol}$
we adopt
\begin{equation}
M_{bol} = -2.54 \log P + 1.98,
\end{equation}
again from LMC carbon Miras but now including some heavily
dust-enshrouded members (Whitelock et al. 2006) and the same
distance modulus. It should be noted that a Galactic zero-point
for this relation ($2.06 \pm 0.24$ (Feast et al. 2006)) is close to the
value used.

With these absolute magnitudes we derive distance moduli of 
$23.09 \pm 0.19$ from $M_{K}$ and $23.10 \pm 0.18$ from $M_{bol}$
where the uncertainties take into account both the scatter about the
PL relations and the uncertainty in the distance modulus of the LMC.
Held et al. (1999) quote moduli of $23.21 \pm 0.08$ from
the horizontal branch (HB) at $V$ and $23.04 \pm 0.07$ from the TRGB at $I$.
Martinez-Delgado (1999) also used the latter method to obtain
a modulus of $23.0\pm 0.1$. Holtzman et al. (2000)
obtained 23.1 from the TRGB and 23.3 from an assumed absolute
magnitude of the HB.
These estimates all agree well and we adopt 23.1 for the galaxy.

M2007 find  a reddening of $E(B-V) = 0.15$ from
a study of the young population of Phoenix. Adopting such a reddening
would decrease the modulus derived from the Mira 
by only about 0.04 mag.
But the
Held et al. (1999) values would be considerably affected; the moduli from the HB
 and from the TRGB would both become 22.81. However,
the M2007 reddening is  likely to apply only to the small young population 
in Phoenix.  Note that if the Phoenix Mira
were an HBB star (as discussed above) it would be expected to be brighter
than the PL relations used here suggest. So the agreement with other
distance moduli noted in the previous paragraph is an indication that
it is not an HBB star.\\

\section{Conclusions}
 By combining our own $JHK_s$ observations with the optical photometry of
Massey et al. (2007) it has been possible to make a rather clean separation
of Phoenix members from field stars.  A clear RGB of an old population 
is found together with a
few highly evolved stars. A Mira variable, almost certainly a carbon star,
with a period of 425 days is present in the galaxy and leads to an estimate
of $23.10 \pm 0.18$ for the distance modulus in agreement with other
estimates.  The kinematics of carbon Miras in our Galaxy (Feast et al. 2006)
suggest an age of $\sim 2\;$Gyr for this star. Since Miras are relatively
short-lived objects this implies a significant population of this age.

The two Da Costa carbon stars have  $M_{K} = -6.2$ or $M_{bol} = -3.8$
(based on an estimate of the bolometric correction from the work of Frogel
et al.(1980)).  These luminosities are consistent with an age $\sim 1$ to a
few Gyr (see e.g. the luminosities of carbon stars in LMC clusters (Frogel
et al. 1990)).  Whilst most of the stars fainter than $K_s \sim 17.2$ are 
found to be members of an old RGB
population, a significant number of them are identified as probably AGB stars
of intermediate age. They are likely to belong to the same population as the carbon
stars. In this connection, we note that a feature
in the colour-magnitude diagram of Holtzman et al. (2000) (their fig. 2),
starting at $V$ or $I$ of $\sim 24.0$, $(V-I) \sim 0$ and sloping to higher
luminosities and redder colours may be a subgiant branch of intermediate age
stars. It is reasonably well fitted by a 1 Gyr isochrone (z = 0.002) from
Girardi et al. (2002).

The status of the non-carbon star Da Costa C2 which is
$\sim 0.3$ mag brighter than the two carbon stars at $K_s$ is uncertain.
It seems most likely to be an AGB star of an old population.
Whether other stars of about the same luminosity and colour to C2 are 
also old AGB stars or carbon stars of an intermediate age population
requires further spectroscopic work.
 
\section*{Acknowledgments}
We are grateful to the IRSF/SIRIUS team, based in  
 Nagoya University, Kyoto University the National Astronomical Observatory
 of Japan, for their support during our observations.
 We also acknowledge our Japanese colleagues,
T. Tanab\'e, Y. Ita, S. Nishiyama, 
R. Kadowaki, A. Ishihara, Y. Haba and J. Hashimoto, for obtaining for us 
some of the images of Phoenix that were used in this investigation.
Dr P. Massey very kindly sent us the Phoenix observations obtained by
himself and his colleagues in advance of publication.

\section*{Appendix}
\renewcommand{\thetable}{A\arabic{table}}
\setcounter{table}{0}
\begin{table*}
\begin{center}
\caption{IRSF data for the non-variable assumed AGB stars more luminous 
than the C stars.}
\begin{tabular}{lccccccccccc}
\hline
No.& $K_s$ & $J-H$ & $H-K_s$ & $J-K_s$ & $V$ & $B-V$ & $V-R$ & $R-I$ & $V-I$ &
$V-K_s$ & $I$ \\
\hline
52 (C2) & 16.59 & 0.85 & 0.09 & 0.94 & 20.335 & 1.720 & 0.929 & 0.796 & 1.725 & 3.75 & 18.610\\
56      & 16.54 & 0.86 & 0.13 & 0.99 & 20.423 & 1.532 & 0.940 & 0.844 & 1.784 & 3.88 & 18.639\\
55      & 16.49 & 0.75 & 0.07 & 0.82 & 19.512 & 1.422 & 0.706 & 0.626 & 1.332 & 3.02 & 18.180\\
60      & 16.73 & 0.82 & 0.14 & 0.96 & 29.266 & 1.510 & 0.843 & 0.730 & 1.573 & 3.54 & 18.693\\
41      & 16.59 & 0.82 & 0.17 & 1.00 & 20.514 & 1.797 & 0.978 & 0.857 & 1.835 & 3.92 & 18.679\\
61      & 16.76 & 0.65 & 0.06 & 0.71 & 19.507 & 1.177 & 0.645 & 0.572 & 1.217 & 2.75 & 18.290\\
57      & 16.56 & 0.87 & 0.16 & 1.04 & 20.649 & 1.892 & 1.003 & 0.868 & 1.871 & 4.09 & 18.778\\
\multicolumn{12}{l}{Carbon stars}\\
87      & 16.90 & 0.81 & 0.15 & 0.96 & 20.636 & 1.718 & 0.907 & 0.725 & 1.632 & 3.74 & 19.004\\
102     & 16.85 & 0.66 & 0.14 & 0.81 & 20.010 & 1.707 & 0.808 & 0.655 & 1.463 & 3.16 & 18.547\\
\hline
\end{tabular}
\end{center}
\end{table*}

\begin{table*}
\begin{center}
\caption{2MASS data for Bessell \& Brett's (1988) dwarfs and giants.}
\begin{tabular}{ccccccccc}
\hline
Sp & $V-K_s$ & $J-H$ & $H-K_s$ & $J-K_s$ &$V-K_s$ & $J-H$ & $H-K_s$ & $J-K_s$\\
& \multicolumn{4}{c}{Dwarfs} & \multicolumn{4}{c}{Giants}\\
\hline
K0 &      &      &      &        &  2.35 & 0.48 & 0.12 & 0.60\\
K1 &      &      &      &        &  2.54 & 0.52 & 0.13 & 0.65\\
K2 &      &      &      &        &  2.74 & 0.57 & 0.14 & 0.71\\
K3 &      &      &      &        &  3.04 & 0.62 & 0.17 & 0.79\\
K4 & 2.67 & 0.52 & 0.13 & 0.65   &  3.30 & 0.67 & 0.18 & 0.85\\
K5 & 2.89 & 0.55 & 0.14 & 0.69   &  3.64 & 0.73 & 0.19 & 0.92\\
K7 & 3.20 & 0.60 & 0.16 & 0.76   &       &      &      &     \\
M0 & 3.69 & 0.64 & 0.19 & 0.83   &  3.89 & 0.77 & 0.22 & 0.99\\
M1 & 3.91 & 0.62 & 0.23 & 0.85   &  4.09 & 0.79 & 0.23 & 1.02\\
M2 & 4.15 & 0.61 & 0.24 & 0.85   &  4.34 & 0.81 & 0.24 & 1.05\\
M3 & 4.60 & 0.56 & 0.28 & 0.84   &  4.68 & 0.84 & 0.26 & 1.10\\
M4 & 5.30 & 0.54 & 0.30 & 0.84   &  5.14 & 0.87 & 0.27 & 1.14\\
M5 & 6.16 & 0.56 & 0.35 & 0.91   &  6.00 & 0.89 & 0.31 & 1.20\\
M6 & 7.34 & 0.60 & 0.40 & 1.00   &  6.88 & 0.90 & 0.33 & 1.23\\
M7 &      &      &      &        &  7.84 & 0.90 & 0.34 & 1.24\\
\hline
\end{tabular}
\end{center}
\end{table*}

\begin{table*}
\begin{center}
\caption{Extreme cool subdwarfs from Gizis (1997); infrared photometry from
2MASS.}
\begin{tabular}{lcccccccccc}
\hline
LHS & $K_s$ & $J-H$ & $H-K_s$ & $J-K_s$ & $V$ & $V-K_s$ & $B-V$ & $V-R$ & $R-I$ & 
$V-I$ \\
\hline
104   &10.412 & 0.525 & 0.154 & 0.679 & 13.78 & 3.37 & 1.34 & 0.81 & 0.91 & 1.72\\
161   &10.995 & 0.516 & 0.203 & 0.719 & 14.75 & 3.75 & 1.55 & 1.01 & 0.96 & 1.98\\
169   &10.819 & 0.474 & 0.193 & 0.667 & 14.13 & 3.31 & 1.45 & 0.91 & 0.76 & 1.72\\
182   &10.519 & 0.428 & 0.150 & 0.578 & 13.42 & 2.90 & 1.57 &      &      &     \\
185   &11.517 & 0.618 & 0.221 & 0.839 & 15.30 & 3.78 & 1.79 & 0.98 & 0.85 & 1.83\\
364   &10.860 & 0.451 & 0.155 & 0.606 & 14.61 & 3.75 & 1.71 & 1.03 & 0.92 & 1.95\\
375   &11.507 & 0.476 & 0.167 & 0.643 & 15.68 & 4.17 & 1.87 & 1.08 & 1.12 & 2.20\\
489   &11.852 & 0.474 & 0.205 & 0.679 & 15.48 & 3.63 & 1.69 & 0.91 & 0.86 & 1.77\\
522   &10.927 & 0.480 & 0.177 & 0.657 & 14.15 & 3.22 & 1.41 & 0.84 & 0.78 & 1.62\\
1970  &13.875 & 0.581 & 0.124 & 0.705 & 17.76 & 3.88 & 1.68 &      &      & 2.09\\
3382  &13.197 & 0.520 & 0.147 & 0.667 & 17.02 & 3.82 & 1.99 & 1.03 & 1.06 & 2.09\\
\hline                                                            
\end{tabular}
\end{center}
\end{table*}


\begin{thebibliography}{}
\bibitem[]{} Aaronson M., Blanco V. M., Cook K. H., Schechter P. L., 1989,
  ApJS, 70, 637
\bibitem[]{} Bahcall J. N., Soneira R. M., 1980, ApJS, 44, 73
\bibitem[]{} Bellazzini M., 2007, arXiv:0711.2016
\bibitem[]{} Bessell M. S., 1990, PASP, 102, 1181
\bibitem[]{} Bessell M. S., Brett J. M., 1988, PASP, 100, 1134
\bibitem[]{} Burstein D., Heiles C., 1982, AJ, 87, 1165
\bibitem[]{} Canterna R., Flower, P. J., 1977, ApJ, 212, 57L 
\bibitem[]{} Carpenter J. M., 2001, AJ, 121, 2851
\bibitem[]{} Da Costa G. S. 1994, in: ESO Conf. Workshop Proc. 49, Dwarf Galaxies: Proc.
ESO/OHP Workshop, eds. G. Meylan \& P. Prugniel (Garching: ESO), 221
\bibitem[]{} Feast M. W., Whitelock P. A., Menzies J.W.,
2006, MNRAS, 369, 791
\bibitem[]{} Frogel J. A., Persson S. E., Cohen J. G., 1980, ApJ, 239, 495
\bibitem[]{} Frogel J. A., Mould J., Blanco V. M., 1990, ApJ, 352, 96
\bibitem[]{} Gallart C., Aparicio A., Freedman W. L., Madore B. F.,
Mart\'i­nez-Delgado, D., Stetson P. B., 2004, AJ, 127, 1486
\bibitem[]{} Girardi L., Bressan A., Bertelli G., Chiosi C., 2000,
A\&ASup, 141, 371
\bibitem[]{} Girardi L., et al., 2002
http://stev.oapd.inaf.it/\~lgirardi/cgi-bin/cmd
\bibitem[]{} Gizis J. E., 1997, AJ, 113, 806
\bibitem[]{} Grebel E., 1999, in: (eds.) Whitelock P. \& Cannon R., The
stellar Content of Local group Galaxies,  IAU Sym. 192, ASP, p. 17
\bibitem[]{} Held E. V., Saviane I., Momany Y., 1999, A\&A, 345, 747	
\bibitem[]{} Holtzman J. A., Smith G. H., Grillmair C., 2000, AJ, 120, 3060
\bibitem[]{} Martinez-Delgado D., Gallart C., Aparicio A., 1999, AJ, 118, 
 862
\bibitem[]{} Massey P., Olsen K. A. G., Hodge P. W., Jacoby G. H.,
McNeill R. T., Smith R. C., Strong S. B., 2007, AJ, 133, 2393 (M2007)
\bibitem[]{} Mateo M., 1998, ARA\&A, 36, 435
\bibitem[]{} Menzies J., Feast M., Tanab\'e T., Whitelock P.A., Nakada Y.,
 2002, MNRAS, 335, 923 (JWM2002)	
\bibitem[]{} Nagashima C. et al., 1999, in Star Formation 1999, 
ed. T. Nakamoto (Nobeyama: Nobeyama Radio Observatory) 397
\bibitem[]{} Nagayama T. et al., 2003, Proc. SPIE 4841, 459
\bibitem[]{} Oosterloo T., Da Costa G. S.,  Staveley-Smith L., 1996, AJ, 112, 1969 
\bibitem[]{} Persson S. E., Murphy D. C., Krzeminski W., Roth M., Rieke M.
J., 1998, AJ, 116, 2457
\bibitem[]{} Pietrinferni A., Cassisi S., Salaris M., Castelli F., 2004, ApJ,
612, 168
\bibitem[]{} Salaris M., Girardi L., 2005, MNRAS, 357, 669
\bibitem[]{} Schuster H. E., West R. M., 1976, A\&A, 49, 129 
\bibitem[]{} St-Germain J., Carignan C., C\^ote S., Oosterloo T., 1999, AJ,
118, 1235
\bibitem[]{} van de Rydt F., Demers S., Kunkel W. E., 1991, AJ, 102, 130
\bibitem[]{} Whitelock P. A., Feast M. W., Marang F., Groenewegen M. A. T.,
 2006, MNRAS, 369, 751
\bibitem[]{} Young L. M., Lo, K. Y., 1997, ApJ, 490, 710 
\bibitem[]{} Young L. M., Skillman E. D., Weisz D. R., Dolphin A. E., 2007,
ApJ, 659, 331
\bibitem[]{}  

\end{thebibliography}
\end{document}